\documentclass[sigconf]{acmart}

\AtBeginDocument{%
  \providecommand\BibTeX{{%
    \normalfont B\kern-0.5em{\scshape i\kern-0.25em b}\kern-0.8em\TeX}}}

\copyrightyear{2019}
\acmYear{2019}
\setcopyright{rightsretained}

\acmConference{VRST '19}{November 12-15, 2019}{Sidney, Australia}
\acmBooktitle{VRST '19, 
November 12-15, 2019, Sidney, Australia}
\acmBooktitle{VRST '19, 
  November 12-15, 2019, Sidney, Australia}

\usepackage{amsmath}
\usepackage{graphicx}
\usepackage{multirow}
\usepackage[caption=false]{subfig}

\begin{document}

\title{Recyglide : A Wearable Multi-modal Stimuli Haptic Display aims to Improve User VR Immersion}
\title{ RecyGlide : A Forearm-worn Multi-modal Haptic Display aimed to Improve User VR Immersion}

\author{Juan Heredia}
\affiliation{%
  \institution{Skolkovo Institute of Science and Technology}
  }
    \email{Juan.Heredia@skoltech.ru}

\author{Jonathan Tirado }
\affiliation{%
  \institution{Skolkovo Institute of Science and Technology}
  }
    \email{Jonathan.Tirado@skoltech.ru}

\author{Vladislav Panov}
\affiliation{%
  \institution{Skolkovo Institute of Science and Technology}
  }
    \email{Vladislav.Panov@skoltech.ru}

\author{Miguel Altamirano Cabrera}
\affiliation{%
  \institution{Skolkovo Institute of Science and Technology}
  }
  \email{Miguel.Altamirano@skoltech.ru}

\author{Kamal Youcef-Toumi}
\affiliation{%
  \institution{Massachusetts Institute of Technology}
  }
    \email{youcef@mit.edu}

\author{Dzmitry Tsetserukou}
\affiliation{%
  \institution{Skolkovo Institute of Science and Technology}
  }
    \email{D.Tsetserukou@skoltech.ru}


\begin{abstract}
Haptic devices have been employed to immerse users in VR environments. In particular, hand and finger haptic devices have been deeply developed. However, this type of devices occlude the hand detection by some tracking systems, or in other tracking systems, it is uncomfortable for the users to wear two hand devices (haptic and tracking device).  We introduce RecyGlide, which is a novel wearable forearm multimodal display at the forearm. The RecyGlide is composed of inverted five-bar linkages and vibration motors. The device provides multimodal tactile feedback such as slippage, a force vector, pressure, and vibration. We tested the discrimination ability of monomodal and multimodal stimuli patterns in the forearm, and confirmed that the multimodal stimuli patterns are more recognizable. This haptic device was used in VR applications, and we proved that it enhances VR experience and makes it more interactive. 
\end{abstract}


\begin{teaserfigure}
  \includegraphics[width=\textwidth]{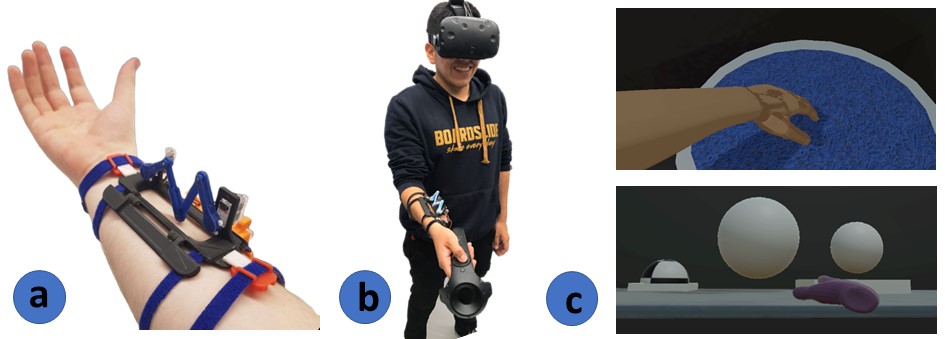}
  \caption{a) Novel  wearable haptic display RecyGlide. b) The user with RecyGlade and Tracking System HTC c) VR Appications }
  \Description{Enjoying the baseball game from the third-base
  seats. Ichiro Suzuki preparing to bat.}
  \label{fig:teaser}
\end{teaserfigure}

\maketitle

\section{Introduction}
Recently, several VR applications have been proposed in many fields: medicine, design, marketing, etc. , and to improve user immersion new methods or instruments are need.  Haptics provides a solution, and enhances user experience using stimuli \cite{Han:2018:HAM:3281505.3281507}. Most of the haptics devices are located in the palm or fingers. However, in some cases the haptic device position is a problem because VR Hand Tracking Systems need a free hand (Leapmotion) or a hand instrument (HTC) to recognize the position.  Therefore, we propose a novel multi-modal haptic display located in the forearm.

Various haptic devices have been developed in the forearm with mono-modal stimulus \cite{Dobbelstein:2018:MSM:3267242.3267249,Moriyama:2018:DWH:3267782.3267795}. However, they have a persistent problem: users have more difficulty perceiving a stimulus in the forearm. The forearm is a not advantageous zone since it does not have as many nerves as the palm or fingertips have.  Consequently, our device produces multi stimuli for improving user perceptions. Furthermore, multimodal stimuli experiments have been employed on the user's hands, where the results show improvements in patterns recognition \cite{10.1007/978-981-13-3194-7_33}.    

RecyGlide is a novel forearm haptic display to provide multimodal stimuli. RecyGlide consists of one inverted five-bar linkage 2-DoF system installed parallel to the radius, which produce a sliding force along the user's forearm. The second stimulus is vibration; two vibration motors are placed in the device's edges (see Fig. 2 (b)). The schematic and 3D representation are shown in Fig 2.

The user study aims to identify the advantages of multimodal stimuli use in comparison with monomodal sensations. The experiment consists of multimodal and monomodal pattern recognition by users. In our hypothesis, two tactile channels could improve the perception of patterns.

\section{Device Development}

$RecyGlide$ provides the sensation of vibration, contact at one point, and sliding at the user's forearm. The location of the haptic contact point is determined using kinematics model of inverted five-bar linkages, inspired by $LinkTouch$ technology \cite{tsetserukou2014}. The second stimulus is generated by two vibration motors located on the extreme sides of the device. The two types of stimuli allow creating different patterns at the forearm. Also, it could be used to interact with the VR environment, e.g., the sensation of submerging in liquids, the felling of an animal movement in the forearm, or information delivery about the relative location.

\begin{figure}[!h]
\centering%
\subfloat[Isometric view of the 3D model.]{%
\label{fig:cylinder}%
\includegraphics[width=0.95\linewidth]{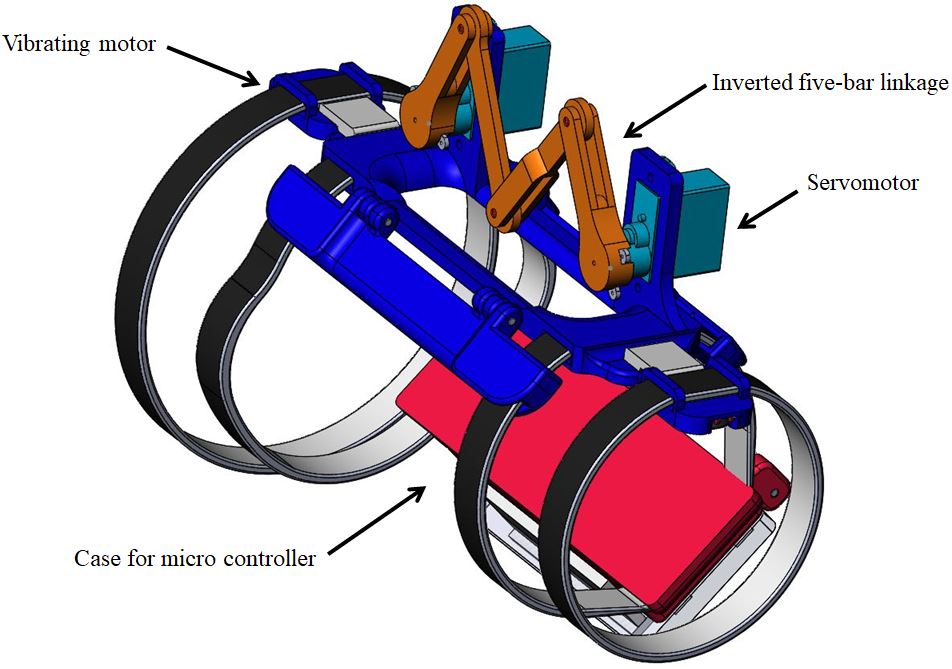}%
} \qquad
\centering%
\subfloat[Front view of the 3D model.]{%
\label{fig:cylinder}%
\includegraphics[width=0.95\linewidth]{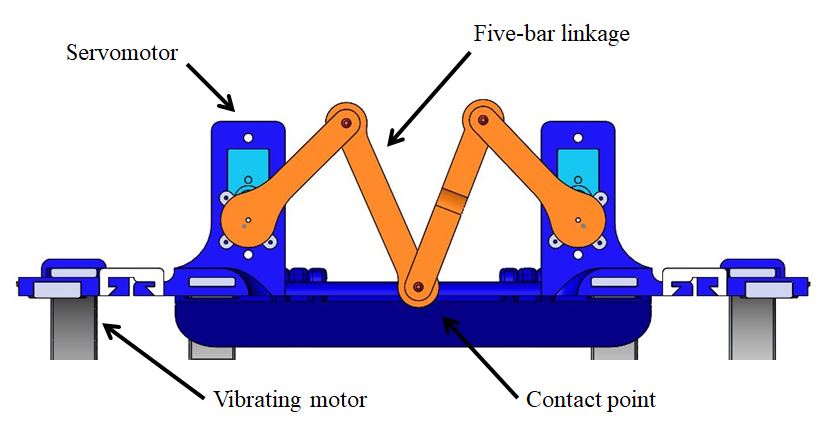}%
} \qquad
\caption{3D CAD model of wearable tactile display.}
\label{fig:linkglidemodel}
\end{figure}

The device has 2-DOF defined by the action of two servomotors. The slippage stimulus is produced by the movement of the motors in the same direction. Conversely, the movement of the motors in the opposite direction generates the force stimulus. The derivation of the previous stimuli creates other stimuli like temperature or pressure. 

$RecyGlide$ location is convenient for hand tracking systems. Commercial tracking systems have different principles to achieve their objective. However, the use of a device in hand decreases the performance. In the case of a visual track, if the haptic device is located in hand, the typical shape of the hand changes producing poor tracking performance. For systems with trackers like HTC,  it is uncomfortable for the user to wear two devices in hand. The device has been designed to adapt ergonomically to the user's forearm, allowing free movement of the hand when working in the virtual reality environment.

The technical characteristics are listed in the Table \ref{char}. The table shows the type of servomotors,and also the weight, the material of the device. 

\begin{table}[h!]
\caption{Technical specification of RecyGlide.}
\label{char}
\centering 
\begin{tabular}{lc}
\hline
Motors & Hitec HS-40\\
Weight & $\ 95 g$\\
Material & PLA and TPU 95A\\
Max. normal force at & \\
 contact point & $2\ N$\\
\hline
\end{tabular}
\end{table}

RecyGlide is electronically composed by an Arduino MKR 1000, servomotors and vibration motors. This Arduino model helps for IoT applications because of its wifi module. The Arduino is in charge of signal generation to the motors and communication wifi TCP/IP to the computer. The device maintains constant wifi communication with the computer throws a python script. A virtual socket transmits the data from python to the VR application in Unity.  

In the experiment, a TCP/ IP server in Arduino was designed. The server provides direct access to the device, avoiding the use of the computer. This utility allows generating apps for cellphones too.  

\section{User Study}

The objective of the following experiment is to analyze the user's perception and recognition of patterns when mononodal and multimodal stimuli are rendered on the forearm, and to determine if the multimodal stimuli increase the user's perception of the contact point position. The first user experience is a contact stimulus over the cutaneous area by the sliding action of the contact point generated by the inverted five-bar linkages device. The second user experience implements stimuli by the combination of vibrating motors and the sliding of the contact point generated by the inverted five-bar linkages device. The results of the experiment will help to understand if the multimodal stimuli improves the perception the position generated by \textit{RecyGlide} device.

\subsection{Experimental Design}

\begin{figure}[h]
  \centering
  \includegraphics[width=1 \linewidth]{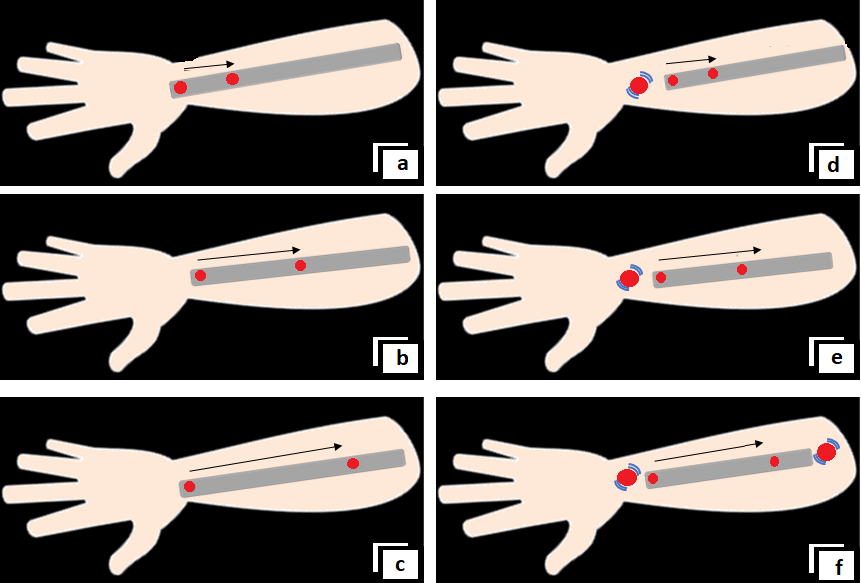}
  \qquad
  \caption{(a)Displacement patterns: (A) Small Distance (SD): progress of 25\% , (B) Medium Distance (MD)progress of 50\% , (C) Large Distance (LG): progress of 75\% , (D) Small Distance with vibration (SDV): progress of 25\%, (E) Medium Distance with vibration (MDV): progress of 50\%, (F) Large Distance with vibration (LDV): progress of 75\%.}
  \label{fig:patterns}
\end{figure}

For the execution of this experiment, a pattern bank of 6 different combinations between displacement and vibratory signals was designed and is shown in the Fig. \ref{fig:patterns}.  In the first group of patterns, A (Small Distance (SD): progress of 25\%), B (Medium Distance (MD): progress of 50\%) and C (Large Distance (LG): progress of 75\%), where monomodal stimuli are delivered; the second group, D (Small Distance with vibration (SDV): progress of 25\%), E (Medium Distance with vibration (MDV): progress of 50\%) and F (Large Distance with vibration (LDV): progress of 75\%), includes multimodal stimuli.The vibration is delivered progressively according to the position of the contact point: while the position of the contact point is nearest to the edge where the vibration motor is located, the vibration frequency is higher; when the contact point is located at the middle distance, the vibration is the same in both vibration motors. The higher frequency used in the vibration motors is equal to $500 Hz$. The sliding speed of the contact point over the skin is constant and has a value of $23 mm/s$.

\subsection{Experimental Setup}
The user was asked to sit in front of a desk, and to wear the $RecyGlide$ device on the right forearm. The device was connected to one Arduino MKR1000. From the python console, the six patterns were sent to the micro-controller throw TCP/IP communication. To reduce the external stimuli, the users wear headphones with white noise. A physical barrier interrupted the vision of the users to their right arm.
Before each section of the experiment, a training session was conducted, where all the patterns were delivered to the users five times. Six participants volunteering complete the experiments, two women and four men, with an average age of 27 years old. Each pattern was delivered on their forearm five times in a random order.

\subsection{Results}
To summarize the data obtained in the experiment, a confusion matrix is tabulated in the Table \ref{confussion}. The diagonal term of the confusion matrix indicates the percentage of correct responses of subjects.

\begin{table}[h!]
\centering
\caption{Confusion Matrix for Patterns Recognition.}
\label{confussion}
\begin{tabular}{|c|c|c|c|c|c|c|c|}
\hline
\multicolumn{2}{|c|}{\textit{\tiny Category}}                                                      
& \multicolumn{6}{c|}{\textit{ \tiny Answers(predicted class)}}                                             
\\ \cline{3-8} 
\multicolumn{2}{|c|}{}                                                                         & \tiny SD           & \tiny MD           & \tiny LD           & \tiny SDV           & \tiny MDV           & \tiny LDV            \\ \hline
\multirow{6}{*}{\tiny \rotatebox{90}{Patterns}} & \tiny Small Distance (SD) & \textbf{77} & 20          & 3           & 0           & 0           & 0            \\ \cline{2-8} 
         & \tiny Medium Distance (MD) & 0           & \textbf{80} & 20          & 0           & 0           & 0            \\ \cline{2-8} 
         & \tiny Large Distance (LD) & 3           & 3           & \textbf{93} & 0           & 0           & 0            \\ \cline{2-8} 
        & \tiny Small Distance Vibration (SDV) & 0           & 0           & 0           & \textbf{97} & 0           & 3            \\ \cline{2-8} 
         & \tiny Medium Distance Vibration (MDV) & 0           & 0           & 0           & 3           & \textbf{93} & 3            \\ \cline{2-8} 
         & \tiny Large Distance Vibration (LDV) & 0           & 0           & 0           & 0           & 0           & \textbf{100} \\ \hline
\end{tabular}
\end{table}

 The results of the experiment revealed that the mean percent correct scores for each subject averaged over all six patterns ranged from 80 to 96.7 percent, with an overall group mean of 90.6 percent of correct answers. Table 1 shows that the distinctive patterns LDV and SDV have higher percentages of recognition 100 and 97, respectively. On the other hand, patterns MD and SD have lower recognition rates of 80 and 77 percent, respectively. For most participants, it was difficult to recognize pattern SD, which usually was confused with pattern MD. Therefore, it is prove the requirement of more distinctive tactile stimuli (vibration) to improve the recognition rate.
 
 The ANOVA results showed a statistically significant difference in the recognition of different patterns (F(5, 30) = 3.2432, p = 0.018532$ <$ 0.05). The paired t-tests showed statistically significant differences between the SD and SDV (p=0.041$<$0.05), and statistical difference between MD and MDV (p=0.025$<$0.05). This results confirm our hypothesis, that the multimodal stimuli improves the perception of the humans in the forearm for short distances. However, the results of paired t-tests between the long distances with and without vibration do not reveal significant differences, thus for long distances perceptions, the multimodal stimuli are not required.

\section{Applications}

\begin{figure}[H]
  \centering
  \includegraphics[width=1 \linewidth]{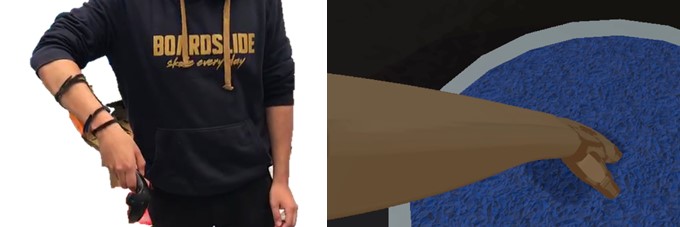}
  \qquad
  \caption{Submerging Hand : moving the contact point the user recognizes the immersion in the liquid.}
  \label{fig:sumerg}
\end{figure}

For the demonstration of RecyGlide, some applications were developed using the game engine Unity 3D. The SubmergingHand application clearly illustrate how the device improves the user immersion in a virtual reality environment. The device provides a sensation of the liquid level by the position of the contact point in the forearm.  The viscosity of the liquid is represented by the normal force applied in the contact point. When the liquid is viscous, the applied force is high and the vibration motors work in a higher frequency. The application is shown in the Fig. \ref{fig:sumerg}. 

The application "Boundaries Recognition" is an example of how our device helps to perceive the VR environment. Habitually VR users go forward the object boundaries and don't respect the physics limits of static objects in the scene, such as walls and tables. RecyGlide informs the boundary collision of the hand tracker by activating the vibration motors and moving the contact point to one of the two sides, which is call collision side. 
 
\begin{figure}[H]
  \centering
  \includegraphics[width=1 \linewidth]{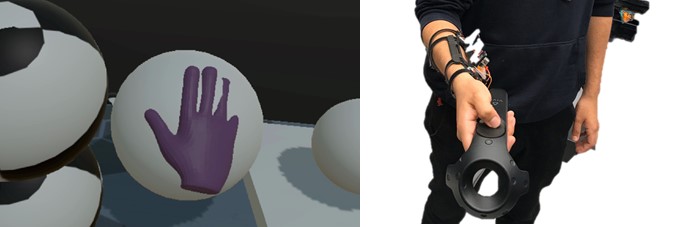}
  \qquad
  \caption{Boundaries Recognition: The display informs when the user has crashed whit an environment boundery}
  \label{fig:patterns}
\end{figure}

These two applications are basic examples, that can be applied in more complex ones. Additionally, the patterns can represent the current state of some variable or environment characteristic; for example, they can communicate the tracker battery status, the selected environment or the distance to an objective, etc.

\subsection{Conclusions}

We proposed a new haptic device in the forearm and made some experiments with it. Due to the results of the experiment, we demonstrated that multimodal stimuli patrons are easily recognizable. Therefore we consider that the device is suitable for communicating some VR messages through the use of patterns.

Though the forearm is not an advantageous area, the device has excellent performance and improves VR realism and the user immersion. The feeling of submersion in liquids can be used in numerous applications such as summing simulators, medical operations, games, etc.  Boundaries collision detection is useful for all kind of VR applications because it is a constant problem in VR environments. 

\bibliographystyle{ACM-Reference-Format}
\bibliography{sample-base}


\begin{thebibliography}{5}


\ifx \showCODEN    \undefined \def \showCODEN     #1{\unskip}     \fi
\ifx \showDOI      \undefined \def \showDOI       #1{#1}\fi
\ifx \showISBNx    \undefined \def \showISBNx     #1{\unskip}     \fi
\ifx \showISBNxiii \undefined \def \showISBNxiii  #1{\unskip}     \fi
\ifx \showISSN     \undefined \def \showISSN      #1{\unskip}     \fi
\ifx \showLCCN     \undefined \def \showLCCN      #1{\unskip}     \fi
\ifx \shownote     \undefined \def \shownote      #1{#1}          \fi
\ifx \showarticletitle \undefined \def \showarticletitle #1{#1}   \fi
\ifx \showURL      \undefined \def \showURL       {\relax}        \fi
\providecommand\bibfield[2]{#2}
\providecommand\bibinfo[2]{#2}
\providecommand\natexlab[1]{#1}
\providecommand\showeprint[2][]{arXiv:#2}

\bibitem[\protect\citeauthoryear{Cabrera and Tsetserukou}{Cabrera and
  Tsetserukou}{2019}]%
        {10.1007/978-981-13-3194-7_33}
\bibfield{author}{\bibinfo{person}{Miguel~Altamirano Cabrera} {and}
  \bibinfo{person}{Dzmitry Tsetserukou}.} \bibinfo{year}{2019}\natexlab{}.
\newblock \showarticletitle{LinkGlide: A Wearable Haptic Display with Inverted
  Five-Bar Linkages for Delivering Multi-contact and Multi-modal Tactile
  Stimuli}. In \bibinfo{booktitle}{\emph{Haptic Interaction}},
  \bibfield{editor}{\bibinfo{person}{Hiroyuki Kajimoto},
  \bibinfo{person}{Dongjun Lee}, \bibinfo{person}{Sang-Youn Kim},
  \bibinfo{person}{Masashi Konyo}, {and} \bibinfo{person}{Ki-Uk Kyung}} (Eds.).
  \bibinfo{publisher}{Springer Singapore}, \bibinfo{address}{Singapore},
  \bibinfo{pages}{149--154}.
\newblock
\showISBNx{978-981-13-3194-7}


\bibitem[\protect\citeauthoryear{Dobbelstein, Stemasov, Besserer, Stenske, and
  Rukzio}{Dobbelstein et~al\mbox{.}}{2018}]%
        {Dobbelstein:2018:MSM:3267242.3267249}
\bibfield{author}{\bibinfo{person}{David Dobbelstein}, \bibinfo{person}{Evgeny
  Stemasov}, \bibinfo{person}{Daniel Besserer}, \bibinfo{person}{Irina
  Stenske}, {and} \bibinfo{person}{Enrico Rukzio}.}
  \bibinfo{year}{2018}\natexlab{}.
\newblock \showarticletitle{Movelet: A Self-actuated Movable Bracelet for
  Positional Haptic Feedback on the User's Forearm}. In
  \bibinfo{booktitle}{\emph{Proceedings of the 2018 ACM International Symposium
  on Wearable Computers}} \emph{(\bibinfo{series}{ISWC '18})}.
  \bibinfo{publisher}{ACM}, \bibinfo{address}{New York, NY, USA},
  \bibinfo{pages}{33--39}.
\newblock
\showISBNx{978-1-4503-5967-2}
\urldef\tempurl%
\url{https://doi.org/10.1145/3267242.3267249}
\showDOI{\tempurl}


\bibitem[\protect\citeauthoryear{Han, Chen, Lee, Wang, Hsieh, Hsiao, Chou, and
  Hung}{Han et~al\mbox{.}}{2018}]%
        {Han:2018:HAM:3281505.3281507}
\bibfield{author}{\bibinfo{person}{Ping-Hsuan Han}, \bibinfo{person}{Yang-Sheng
  Chen}, \bibinfo{person}{Kong-Chang Lee}, \bibinfo{person}{Hao-Cheng Wang},
  \bibinfo{person}{Chiao-En Hsieh}, \bibinfo{person}{Jui-Chun Hsiao},
  \bibinfo{person}{Chien-Hsing Chou}, {and} \bibinfo{person}{Yi-Ping Hung}.}
  \bibinfo{year}{2018}\natexlab{}.
\newblock \showarticletitle{Haptic Around: Multiple Tactile Sensations for
  Immersive Environment and Interaction in Virtual Reality}. In
  \bibinfo{booktitle}{\emph{Proceedings of the 24th ACM Symposium on Virtual
  Reality Software and Technology}} \emph{(\bibinfo{series}{VRST '18})}.
  \bibinfo{publisher}{ACM}, \bibinfo{address}{New York, NY, USA}, Article
  \bibinfo{articleno}{35}, \bibinfo{numpages}{10}~pages.
\newblock
\showISBNx{978-1-4503-6086-9}
\urldef\tempurl%
\url{https://doi.org/10.1145/3281505.3281507}
\showDOI{\tempurl}


\bibitem[\protect\citeauthoryear{Moriyama, Nakamura, and Kajimoto}{Moriyama
  et~al\mbox{.}}{2018}]%
        {Moriyama:2018:DWH:3267782.3267795}
\bibfield{author}{\bibinfo{person}{Taha Moriyama}, \bibinfo{person}{Takuto
  Nakamura}, {and} \bibinfo{person}{Hiroyuki Kajimoto}.}
  \bibinfo{year}{2018}\natexlab{}.
\newblock \showarticletitle{Development of a Wearable Haptic Device That
  Presents the Haptic Sensation Corresponding to Three Fingers on the Forearm}.
  In \bibinfo{booktitle}{\emph{Proceedings of the Symposium on Spatial User
  Interaction}} \emph{(\bibinfo{series}{SUI '18})}. \bibinfo{publisher}{ACM},
  \bibinfo{address}{New York, NY, USA}, \bibinfo{pages}{158--162}.
\newblock
\showISBNx{978-1-4503-5708-1}
\urldef\tempurl%
\url{https://doi.org/10.1145/3267782.3267795}
\showDOI{\tempurl}


\bibitem[\protect\citeauthoryear{Tsetserukou, Hosokawa, and
  Terashima}{Tsetserukou et~al\mbox{.}}{2014}]%
        {tsetserukou2014}
\bibfield{author}{\bibinfo{person}{D. Tsetserukou}, \bibinfo{person}{S.
  Hosokawa}, {and} \bibinfo{person}{K. Terashima}.}
  \bibinfo{year}{2014}\natexlab{}.
\newblock \showarticletitle{LinkTouch: A wearable haptic device with five-bar
  linkage mechanism for presentation of two-DOF force feedback at the
  fingerpad}. In \bibinfo{booktitle}{\emph{2014 IEEE Haptics Symposium
  (HAPTICS)}}. \bibinfo{pages}{307--312}.
\newblock
\showISSN{2324-7355}
\urldef\tempurl%
\url{https://doi.org/10.1109/HAPTICS.2014.6775473}
\showDOI{\tempurl}


\end{thebibliography}


\end{document}